\documentclass[aps,prl,twocolumn,notitlepage,superscriptaddress]{revtex4-1}

\usepackage[utf8]{inputenc}
\usepackage{amsmath,amsfonts,amssymb}
\usepackage{graphicx}

\usepackage{hyperref}
\usepackage[usenames,dvipsnames]{xcolor}

\usepackage{tikz}
\hypersetup{colorlinks=true, linkcolor=BrickRed, urlcolor=blue!50!black, citecolor=blue!50!black}

\newcommand\diff{\mathrm{d}}

\renewcommand{\vec}[1]{\mathbf{#1}}
\renewcommand{\phi}[0]{\varphi}

\usepackage[normalem]{ulem}

\begin{document}

\title{Persistent anti-correlations  in Brownian dynamics simulations of dense colloidal suspensions revealed by  noise suppression}

\author{Suvendu Mandal}
\affiliation{Institut f\"ur Theoretische Physik II: Weiche Materie, Heinrich-Heine-Universit\"at D\"usseldorf, Universit\"atsstra{\ss}e 1, 40225 D\"usseldorf, Germany}
\author{Lukas Schrack}
\affiliation{Institut f\"ur Theoretische Physik, Universit\"at Innsbruck, Technikerstra{\ss}e 21A, A-6020 Innsbruck, Austria}
\author{Hartmut L\"owen}
\affiliation{Institut f\"ur Theoretische Physik II: Weiche Materie, Heinrich-Heine-Universit\"at D\"usseldorf, Universit\"atsstra{\ss}e 1, 40225 D\"usseldorf, Germany}
\author{Matthias Sperl}
\affiliation{Institut f\"ur Materialphysik im Weltraum, Deutsches Zentrum f\"ur Luft- und Raumfahrt,  51170 K\"oln, Germany}
\affiliation{Institut f\"ur Theoretische Physik, Universit\"at zu K\"oln, Z\"ulpicher Stra{\ss}e 77, 50937 K\"oln, Germany}
\author{Thomas Franosch}
\affiliation{Institut f\"ur Theoretische Physik, Universit\"at Innsbruck, Technikerstra{\ss}e 21A, A-6020 Innsbruck, Austria}

\begin{abstract}
Transport properties  
of a hard-sphere colloidal fluid are investigated by Brownian dynamics simulations. We implement a novel algorithm for the time-dependent velocity-autocorrelation function (VACF) essentially eliminating the noise of the bare random motion. 
The measured VACF reveals  persistent  anti-correlations manifested by a negative
 algebraic power-law tail $t^{-5/2}$ at all densities.  
At small packing fractions the simulations fully agree with the analytic low-density prediction, yet  the amplitude of the tail becomes dramatically suppressed   as the  packing fraction is increased.  
The mode-coupling theory of the glass transition provides a qualitative explanation for the strong variation   in terms of the static compressibility as well as the  slowing down of the structural relaxation.

\end{abstract}

\date{\today}
\maketitle

\paragraph{Introduction.--}
In a fluid the velocity autocorrelation function (VACF) in equilibrium encodes the  self-diffusion coefficient as  its time-integral, similar
 Green-Kubo relations exist for all transport coefficients such as viscosity or heat conductivity~\cite{Hansen:Theory_of_Simple_Liquids}. For underlying  Newtonian dynamics it is well established since the late  1960' by the pioneering simulations of Alder and Wainwright~\cite{Alder:PRL:1967, *Alder:PRA_1:1970}, exact theoretical results~\cite{Ernst:PRL_25:1970,Dorfman:PRL_25:1970}, and experiments~\cite{Lukic:PRL_95:2005,Jeney:PRL_100:2008,Franosch:Nat_478:2011,Huang:Nature_Physics:2011} that the VACF and other relevant correlation functions display 
an algebraic power-law decay $t^{-3/2}$ in 3d. Such tails then imply a non-analytic behavior for the frequency-dependent transport coefficients. The origin of these persistent correlations is traced back to the slow diffusion of conserved transverse momentum~\cite{Alder:PRL:1967, *Alder:PRA_1:1970, Hansen:Theory_of_Simple_Liquids, Lesnicki:PRL_116:2016, Peng:PRE_94:2016}. Similar tails also arise due to the presence of boundaries \cite{Felderhof:JPCB_109:2005,Jeney:PRL_100:2008,Huang:NatComm_6:2015} or in the presence of disorder~\cite{Frenkel:PLA_121:1987,Williams:PRL_96_8:2006,Hoefling:PRL_98:2007} or even driven granular systems~\cite{Fiege:PRL_102:2009,*Gholami:PRE_84:2011}.

For a colloidal suspension, velocity is not an observable anymore, rather it fluctuates without bounds, momentum is lost incessantly due to friction and gained by thermal noise. Nevertheless, a VACF $Z(t)$ can be defined formally by
\begin{align}\label{eq:VACF_MSD}
 Z(t) := \frac{1}{6} \frac{\diff^2 }{\diff t^2} \langle [\vec{R}(t)-\vec{R}(0)]^2 \rangle , \qquad t> 0,
\end{align} 
thereby extending the connection with the mean-square displacement also to the colloidal case.  
Recently, there has been experimental progress to monitor with high precision  the VACF of an isolated colloid in a solvent held by an optical trap~\cite{Lukic:PRL_95:2005,Jeney:PRL_100:2008,Franosch:Nat_478:2011,Huang:Nature_Physics:2011} and confirm that the coupling to the solvent gives rise to long-time tails as above and colored noise as predicted theoretically~\cite{Zwanzig:PRA_2:1970,Widom:PRA:1971}. 

In Brownian dynamics (BD) the solvent is treated only implicitly by Gaussian white noise and friction, correspondingly these hydrodynamic tails due to momentum conservation do not occur. Yet,  exact low-density expansions~\cite{Hanna:JPAMG_14:1981, * Hanna:PhysA_111:1982,Ackerson:JCP_76:1982,Felderhof:PhysA_121:1983, * Felderhof:PhysA_122:1983} developed in the early 1980'  of the many-body Smoluchowski equation in $3$D 
predict a similar long-time behavior for the VACF, although more rapidly decaying $t^{-5/2}$ and with negative prefactor. The tails arise due to particle conservation and reflect the repeated encounters with the same scatterer.
Roughly speaking, the particle 'remembers' that a second colloid is blocking the way in the relative motion,  the constraint fading away only slowly by diffusion. Generally, these tails  appear as universal feature of strongly interacting particle systems lacking momentum conservation.  
In contrast to the hydrodynamic tails, there appears to be only rudimentary data analysis~\cite{Cichocki:PRA_44:1991} of BD simulation results to corroborate the persistent tails. The difficulty in obtaining accurate results in 
BD simulations is that the mean-square displacement is dominated by the white noise keeping the dynamics in equilibrium. For dilute systems the interactions leading to deviations from conventional diffusion are rare events and get buried completely in the noise. For higher packing fractions there is a significant suppression of the diffusive motion, yet the persistent tails (if they prevail beyond the low-density regime) are again hard to extract from noisy data. 

In this Letter we present BD simulation data for the VACF over several orders of magnitude in time and amplitude thereby confirming the existence of such long-time tails for the first time. The key ingredient is an adaption of an algorithm originally introduced by Frenkel~\cite{Frenkel:PLA_121:1987} for a single particle on a lattice. At the lowest packing fractions $\varphi$ the data are fully described by the low-density expansion, in particular we reproduce 
the short-time divergence as well as the predicted long-time tail. We show that the mode-coupling theory (MCT) of the glass transition for colloids also yields a long-time tail provided the long-wavelength dynamics is properly resolved. The trends in the amplitude of the tail in the simulation are rationalized within MCT. Furthermore, we elaborate the MCT prediction for the VACF in the vicinity of the glass transition.

\paragraph{Model and Simulation.--} To unravel the VACF in detail, we  investigate a hard-sphere colloidal fluid with particles of identical diameters $\sigma$ and short-time diffusion coefficient $D_0$ at packing fraction $\phi = n \sigma^3 \pi/6$~\cite{Hansen:Theory_of_Simple_Liquids}. Correspondingly, $t_0 = \sigma^2/D_0$ sets the basic unit of time, 
$Z(t) \sigma^2 /D_0^2$
 is the dimensionless VACF.  
We rely on  event-driven Brownian dynamics simulations~\cite{Scala:JCP_126:2007,Strating:PRE_59:1999,Tao:JCP_124:2006,Leitmann:PRF_3:2018}
for $1000$ particles using a fixed Brownian time step and evolve the particles ballistically including collisions. 

Yet, for low densities most of the time no collisions occur and the dynamics is dominated by the noise of the free Brownian motion.   
Motivated by Frenkel~\cite{Frenkel:PLA_121:1987}, we propose a novel method, which generates two trajectories for an \emph{identical} noise history,  one with collisions and another for free Brownian motion ignoring  interactions. Accordingly, we split the total displacement of a particle   into two contributions
\begin{align}
\Delta \vec{R}(t)  
& = \Delta \vec{R}_{0}(t) + \delta \vec{R}(t),  
\label{eq:displacement}
\end{align}
where  $\Delta \vec{R}_{0}(t)$ represents the random displacement of the non-interacting trajectory. The difference   $\delta \vec{R}(t)$ is then  the collision-induced displacement. 
 Then the mean-square displacement evaluates to 
\begin{align}\label{eq:MSD_noise_suppression}
\langle \Delta \vec{R}(t)^2 \rangle  
&= 6D_{0}t  - \langle \delta \vec{R}(t)^2 \rangle + 2 \langle \Delta \vec{R}(t) \cdot \delta \vec{R}(t) \rangle. 
\end{align}

\begin{figure}[h]
\includegraphics[width=\linewidth]{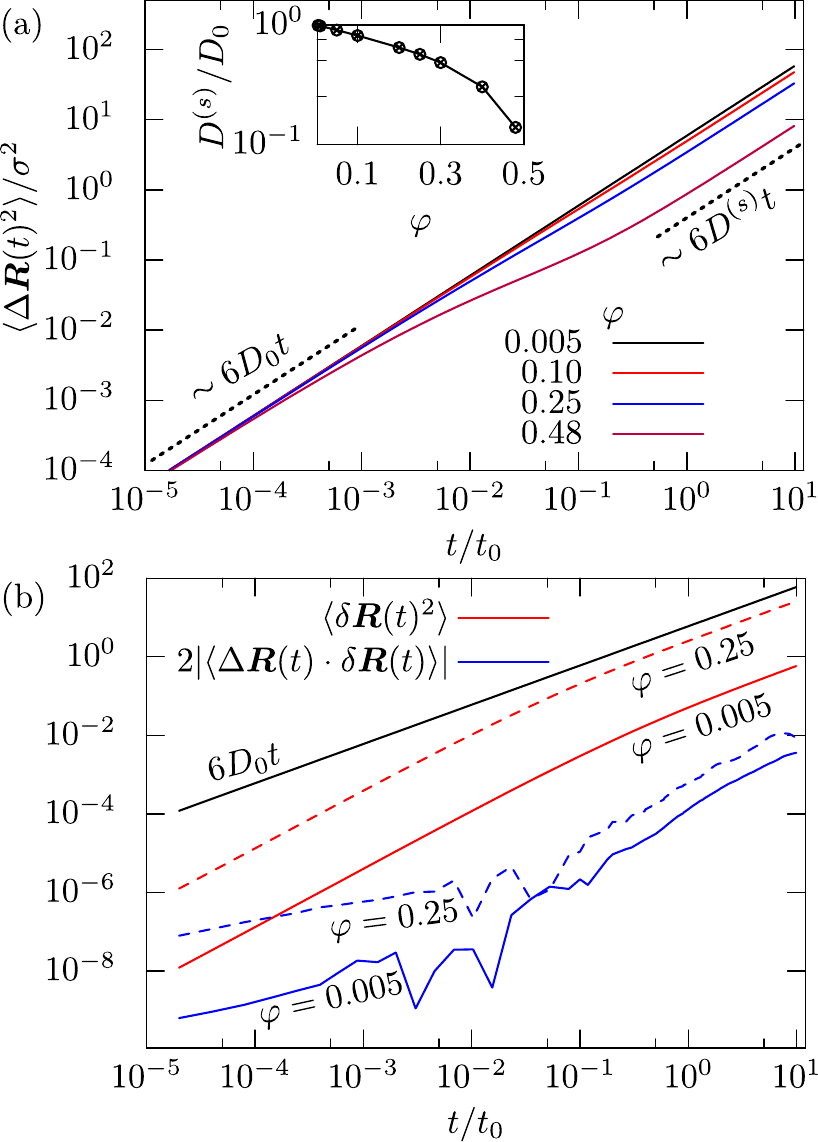} \\
	\caption{\label{fig:Simulation_MSD}
(a) Mean-square displacement $\langle \Delta \vec{R}(t)^2 \rangle $ of monodisperse Brownian hard spheres for  various packing fractions $\varphi$. Inset: Long-time diffusion coefficients $D^{(s)}$ vs packing fraction. (b) Each term in Eq.~\eqref{eq:MSD_noise_suppression} for packing fractions $\varphi=0.005$ (solid lines) and $\varphi=0.25$ (dashed lines).
}
\end{figure}

\begin{figure}                                                                                           
\includegraphics{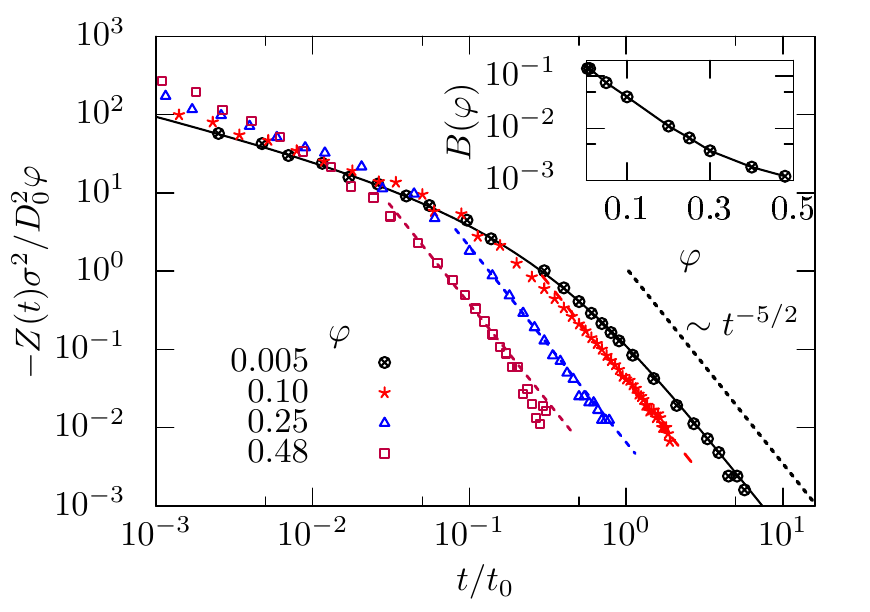}  \\
\caption{\label{fig:Simulation_VACF}                                                                     
Log-log plot of the reduced VACF of monodisperse Brownian hard spheres for increasing packing fraction $\varphi$. Symbols represent simulation data. The black solid line is the exact low-density expansion, Eq.~\eqref{eq:VACF_low_density}. The colored dashed lines are fits to the power-law tails.
The dashed black line labeled $t^{-5/2}$ is added as  guide to the eye. Inset: Prefactor of the tail vs. packing fraction. 
}
\end{figure}

The mean-square displacement (MSD) is evaluated  for 
packing fractions ranging from  the dilute regime, $\varphi = 0.005$, to just below the freezing transition,  $\varphi =0.48$, see Fig.~\ref{fig:Simulation_MSD}. Both the short-time diffusive motion $\langle \Delta \vec{R}(t)^2 \rangle = 6 D_0 t, t\to 0$,
as well as the long-time diffusion  $\langle \Delta \vec{R}(t)^2 \rangle = 6 D^{(s)} t, t\to \infty$  is properly resolved. From the data we extract the long-time self-diffusion coefficient $D^{(s)}$ and observe a slowing down by a factor of $\approx 7$ with respect to the free motion. The crossover regime extends over 2 decades for the highest densities but appears to be a featureless smooth interpolation. 

Clearly, for low densities the correction terms become small relative to the bare diffusion term by construction. Yet, the simulations reveal that the cross-correlation term $2 \langle \Delta \vec{R}(t) \cdot \delta \vec{R}(t) \rangle$ in the MSD is at least by 2 orders of magnitude smaller  than the correction term $\langle \delta \vec{R}(t)^2 \rangle$, see Fig.~\ref{fig:Simulation_MSD}(b). We have checked that this hierarchy of contributions persists to all densities. This observation suggests that we can ignore the cross-correlation completely. Then the collision-induced mean-square displacement captures the suppression of diffusion, reflecting that interaction can only slow down the MSD.  

Upon taking derivatives in Eq.~\eqref{eq:MSD_noise_suppression}, the bare diffusion term drops out for $t>0$ while it formally yields a contribution $6 D_0 \delta(t)$ at the time origin. 
The remaining terms can be differentiated numerically and yield high-accuracy data for the VACF. Note, that in the low-density regime $\delta \vec{R}(t)$ evaluates to zero most of the time, since few collisions occur. Using the observation that the cross-correlation can be ignored, a noise suppression by more than 2 orders of magnitude is achieved at low densities, see Supplementary Material~\cite{supplement_colloidal_vacf}.

The computed  VACF reveal non-trivial correlations beyond the crossover regime, see Fig.~\ref{fig:Simulation_VACF}. Our data cover 5 decades in time and more than 5 orders of magnitude in signal. 
The data clearly display an anti-correlated long-time tail  
\begin{equation}\label{eq:VACF_tail_prefactor} 
Z(t) \simeq -\varphi B(\varphi) \frac{D_0^2}{\sigma^2}  \left(\frac{t}{t_0}\right)^{-5/2}, \qquad  t\to \infty ,
\end{equation}
for all packing fractions 
with the anticipated exponent $-5/2$ and dimensionless  prefactor $\varphi B(\varphi)$. Since a non-trivial VACF arises only in the interacting system, we have included $\varphi$ explicitly in the prefactor. Our simulation results show that $B(\varphi\to 0) \approx 0.14 $ saturates for small packing fractions.  
For the lowest densities we compare to the exact analytical results~\cite{Hanna:JPAMG_14:1981,  * Hanna:PhysA_111:1982,Ackerson:JCP_76:1982,Felderhof:PhysA_121:1983, * Felderhof:PhysA_122:1983} 
\begin{align}\label{eq:VACF_low_density}
 Z(t) = - 8\varphi \frac{D_0^2}{\sigma^2} \Big\{& \sqrt{\frac{t_0}{2\pi t}} - \cos(4 t/t_0) \left[ 1 - 2 S( \sqrt{8t/\pi t_0}) \right] \nonumber \\
 &+ \sin(4 t/t_0) \left[1- 2 C(\sqrt{8 t/\pi t_0}) \right]\Big\}, 
\end{align}
for the first-order in the low-density expansion. Here $S(\cdot), C(\cdot)$ denote the Fresnel integrals~\cite{Olver:2010:NHMF, * NIST:DLMF}. Our data nicely follow the theoretical prediction for $\varphi \lesssim 0.01$,  
in particular, they exhibit the long-time tail with exact prefactor $B(\varphi\to 0) = 3/(8 \sqrt{2\pi})$. 
For larger packing fractions, the prefactor $B(\varphi)$ displays a strong density dependence   beyond the low-density prediction, Fig.~\ref{fig:Simulation_VACF}[Inset].  We find that the amplitude is suppressed by a factor of  $\approx 110$ upon approaching the freezing transition.

All simulation data also display a divergent short-time behavior  $Z(t) \sim - t^{-1/2}, t\to 0$ which is a peculiarity of the hard-sphere interaction~\cite{Cichocki:PhysA_204:1994,Banchio:JCP_113:2000}, the discussion is deferred to the Supplementary Material~\cite{supplement_colloidal_vacf}.

\begin{figure}[htp]
\includegraphics{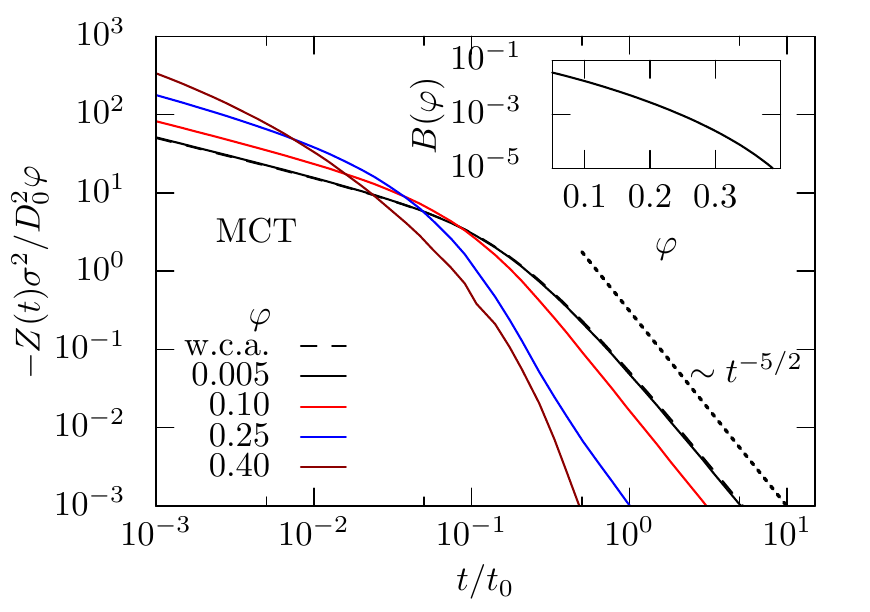}
\caption{\label{fig:MCT_VACF}
Mode-coupling-theory results for the reduced VACF $Z(t)/\varphi$  for  packing fraction $\varphi$ with  Percus-Yevick closure  such that the diffusion coefficients match the simulation results. 
The short-dashed line indicates a power law $t^{-5/2}$ and serves as a guide to the eye. The long-dashed line corresponds to the weak-coupling approximation (w.c.a.) and is virtually on top of the MCT result for the lowest packing fraction. Inset: Prefactor of the tail vs. packing fraction. }   
\end{figure}

\paragraph{Mode-coupling theory.--} Numerical solutions of the colloidal MCT equations~\cite{Szamel:PRA_44:1991,Goetze:Complex_Dynamics} for the self-motion  and, in particular the  MSD,
 have been developed earlier~\cite{Fuchs:PRE_58:1998} and successfully compared to experiments~\cite{Megen:PRE_58:1998,Sperl:PRE_71:2005,Voigtmann:PRE_68:2003,Williams:PRE_64:2001} and simulations~\cite{Voigtmann:PRE_70_2004}. Thus, in principle the VACF can be obtained by taking derivatives as in Eq.~\eqref{eq:VACF_MSD}. Yet, so far MCT equations relied on equidistant grids in wavenumber space and cannot properly resolve long-wavelength phenomena arising from a continuum of wavenumbers. 

Here we rely on the Zwanzig-Mori procedure~\cite{Szamel:PRA_44:1991,Cichocki:Physica_A_141:1987,Kawasaki:Physica_A_215:1995} for an  exact equation of motion for the VACF in terms of the irreducible memory kernel $\zeta^{(s)}(t)$  (the autocorrelation function of the fluctuating force with projected dynamics)   
\begin{align}\label{eq:VACF_eom}
 Z(t) + D_0^2 \beta^2 \zeta^{(s)}(t) + D_0 \beta^2 \int_0^t \zeta^{(s)}(t-t') Z(t') \diff t' =0 , 
\end{align}
for $t>0$. Within MCT~\cite{Goetze:Complex_Dynamics} the memory kernel is connected to the self-intermediate scattering functions of the collective $S(k,t)$ and self-motion $S^{(s)}(k,t)$ by an integral over all wavenumbers
\begin{align}\label{eq:MCT_force_kernel}
 \beta^2 \zeta^{(s)}(t)  = \frac{n}{6\pi^2 } \int_0^\infty\diff k \,  k^4  c(k)^2 S(k,t) S^{(s)}(k,t),
\end{align}
where $c(k)$ is the direct correlation function~\cite{Hansen:Theory_of_Simple_Liquids}. The intermediate scattering functions $S(k,t), S^{(s)}(k,t)$ are required 
as input for the memory kernel $\zeta^{(s)}(t)$. Then the VACF follows by a numerical solution of Eq.~\eqref{eq:VACF_eom}.
We  solve numerically the standard MCT equations, yet to resolve the long-wavelength dynamics, we rely  on a grid with logarithmic spacing, see Supplementary Material~\cite{supplement_colloidal_vacf} for details.


The numerical results for the VACF within MCT display a long-time tail for all packing fractions,  see  Fig.~\ref{fig:MCT_VACF}. It is well-known that MCT overestimates the slowing-down or the structural relaxation. For comparison with the simulation results, the packing fractions in MCT have been chosen to match the suppression of diffusion obtained from simulation results. 
The overall shape of the VACF compares favorably to the simulation results, in particular, we reproduce the strong density dependence of the prefactor of the long-time tail. Nevertheless, it appears that the suppression of the tail is even more drastic in MCT than in simulations, see  inset of Fig.~\ref{fig:MCT_VACF}. The MCT results show a similar short-time divergence $t^{-1/2}$ for $t\to 0$ as the simulation data and the exact low-density result. 

To gain further insight into how MCT encodes the tail we analyze the behavior of the equations analytically for low densities and  for long and short times.  
In the low-density regime $c(k) \mapsto 4 \pi [ \sin(k \sigma)- k \sigma \cos(k\sigma)  ]/k^3$, 
the force kernel $\zeta^{(s)}(t)$, Eq.~\eqref{eq:MCT_force_kernel},  simplifies and reproduces the weak-coupling approximation~\cite{Marqusee:JCP_73:1980,Dhont:Introduction_to_Dynamics_of_Colloids,Naegele:PhysRep:1996,Naegele:JCP_108:1998} (essentially second order perturbation in the interaction). Here the intermediate scattering functions have to be evaluated for the non-interacting system $S(k,t) \mapsto  \exp(- D_0 k^2 t),  S^{(s)}(k,t) \mapsto  \exp(- D_0 k^2 t)$.
Then, for low densities one finds  $Z(t) = - D_0^2 \beta^2 \zeta^{(s)}(t)$ from Eq.~\eqref{eq:VACF_eom} for all times. The analytical result of the weak-coupling approximation is included in Fig.~\ref{fig:MCT_VACF} and coincides with the numerical MCT solution for $\varphi \leq 0.005$. Interestingly, for the rather low packing fraction $\varphi = 0.10$,  MCT yields already a  long-time tail in $Z(t)/\varphi$ suppressed by a factor of $5$, 
while in our simulations, $Z(t)/\varphi$ it  is suppressed by a factor of $4$. 

Mode-coupling theory provides an explanation for the origin of the tails similar to the classic hydrodynamic tails~\cite{Ernst:PRL_25:1970,Dorfman:PRL_25:1970} which are due to transverse momentum conservation. In our case the coupling of the collective and self-intermediate scattering function in the force, Eq.~\eqref{eq:MCT_force_kernel}, yields a slowly decaying contribution for long wavelengths.  
For long times the integral   is dominated by small wavenumbers where  the intermediate scattering functions  $S(k,t) \simeq S(0) \exp(-D_0  k^2 t/S(0)), S^{(s)}(k,t) \simeq \exp(-D^{(s)} k^2 t)$ approach diffusive motion. Here 
$D^{(s)}$ is again the long-time self-diffusion coefficient and  $D_0/S(0)$ the collective diffusion coefficient (which is time-independent as a consequence of Newton's third law~\cite{Dhont:Introduction_to_Dynamics_of_Colloids}). Then the MCT approximation yields a long-time tail 
$\zeta^{(s)}(t) \sim t^{-5/2}$ with positive prefactor. 
The long-time behavior of the VACF $Z(t) \simeq - (D^{(s)} )^2 \beta^2 \zeta^{(s)}(t)$ follows 
from the equation of motion, Eq.~\eqref{eq:VACF_eom}, using Tauber theorems. Collecting the prefactors yields the asymptotic long-time behavior 
\begin{align}\label{eq:VACF_tail_MCT}
\text{MCT: } Z(t) \simeq  - (D^{(s)})^2 \frac{S(0) n c(0)^2/16\pi^{3/2} }{[D^{(s)} + D_0/S(0) ]^{5/2}} t^{-5/2} \, .
\end{align}
In particular, this yields a prediction for the low-density behavior  
with dimensionless amplitude $B(\varphi) = (1/6 \sqrt{2\pi})$  
 consistent with the weak-coupling result~\cite{Dhont:Introduction_to_Dynamics_of_Colloids}. Thus, the amplitude in  weak-coupling is by a factor of $4/9$ smaller than the exact low-density expansion due to repeated encounters with the same scatterer.

It is interesting to ask how the long-time anomaly evolves with increasing density. At moderate packing fractions the fluid is barely compressible, $S(0) \ll 1$ 
(while  $n c(0) = 1-1/S(0) \approx -1/S(0)$ via the Ornstein-Zernike relation~\cite{Hansen:Theory_of_Simple_Liquids}), hence the self-diffusion coefficient can be ignored in the denominator of Eq.~\eqref{eq:VACF_tail_MCT} and
the prefactor displays a strong $\sim S(0)^{3/2}$ dependence by mere compressibility effects. Upon changing the packing fraction from $\varphi=0.005$ to $\varphi=0.40$  
the static structure factor $S(0)$ is suppressed by a factor of $24$. Approaching the glass transition, the self-diffusion coefficient $D^{(s)}$ singularly goes to zero such that the tail in Eq.~\eqref{eq:VACF_tail_MCT} becomes even more suppressed. At the same time the structural relaxation diverges and an intermediate window between the short-time anomaly and the long-time tail should open. 
Currently it appears to be unfeasible to test these predictions by simulations.

\paragraph{Summary and Conclusion.--} We have measured  the VACF of an interacting colloidal system in Brownian dynamics simulations and observed an anti-correlated algebraic decay for long times. These underlying  persistent correlations are  masked in the MSD since the diffusive increase usually dominates. Therefore we have elaborated a novel algorithm which is sensitive only to the collisions, thereby enhancing the signal-to-noise ratio
at least by an order of magnitude. The amplitude of the tail decreases by several orders of magnitude as the packing fraction is increased, which is qualitatively reproduced by MCT. The origin of the tail is also rationalized by MCT as a result of coupling of two slow diffusive modes. 

While the low-density expansion is valid only up to packing fractions $\varphi \lesssim 0.01$,  MCT provides a prediction for all densities, in particular, it predicts the strong suppression of the amplitude by static compressibility effects as well as by the slowing down of self-diffusion. MCT also generalizes the weak-coupling approximation, where the direct  correlation function is replaced by the bare interaction potential, $c(k) \mapsto -\beta u(k)$. This replacement arises also in a factorization of Gaussian fluctuations \cite{Zaccarelli:EPL_55:2001,Wu:PRE_67:2003} but can be avoided in diagrammatic approaches~\cite{Szamel:JCP_127:2007, *Szamel:ProgTEP_1:2013}. Generally, the tails should also be present in modified  MCT approaches~\cite{Yeomans-Reyna:PRE_76:2007,Krakoviack:SoftMatter_13:2017,*Krakoviack:PRE_82:2010,*Krakoviack:PRL_94_2005,Wu:PRL_95:2005, Janssen:PRL_115:2015,*Janssen:PRE_90:2014,Nandi:PRL_119:2017,Contreras:JCP_139:2013,Banchio:JCP_148:2018}.   
Similar persistent correlations are anticipated also for the time-dependent stress-stress correlation functions which determines the frequency-dependent shear modulus also encoded in the MCT approach.

In Brownian dynamics the solvent exerts friction only on the single-colloid level  while  hydrodynamic interactions (HI) can be accounted for in Stokesian dynamics simulations~\cite{Brady:AnnRevFluid:1988}. 
The low-density prediction~\cite{Hanna:JPAMG_14:1981, * Hanna:PhysA_111:1982,Ackerson:JCP_76:1982,Felderhof:PhysA_121:1983, *Felderhof:PhysA_122:1983} has been generalized 
to incorporate HI and display the same long-time tail $ t^{-5/2}$, albeit with a somewhat corrected prefactor~\cite{Jones:PhysicaA:1982}. Similarly, MCT  including HI~\cite{Naegele:PhysRep:1996,Naegele:JCP_108:1998} merely modifes the vertex thereby affecting only the prefactor of the tail. For high densities HI are believed not to be crucial to understand the slow structural relaxation~\cite{Hunter:RPP_75:2012}. The tail appears also in the case of 
 quenched disorder~\cite{Krakoviack:SoftMatter_13:2017,*Krakoviack:PRE_82:2010,*Krakoviack:PRL_94_2005,Pellicane:PRE_69:2004}, such as in the Lorentz gas explored by a Brownian tracer as predicted in a low-density expansion~\cite{Franosch:CP_375:2010} and was confirmed in Brownian dynamics simulations in 2d~\cite{Bauer:EPJ_189:2010}.
Therefore, one anticipates that the emergence of the tail should be a universal feature for any dynamics conserving only the particle number~\cite{vanBeijeren:RMP_54:1982}.


Noise-suppression algorithms relying on identical noise histories have been introduced before, mainly in non-equilibrium Brownian dynamics~\cite{Oettinger:Macromolecules:1994, * Melchior_JCP_105:1996, *Wagner:JRheo_41:1997,Evans:Statistical_Mechanics:2008}. There, the equilibrium fluctuations are subtracted to enhance the signal for the average response for small driving which otherwise is dominated by fluctuations.   
The method proposed in our work   is somewhat different, rather it  addresses the interactions of the particles and remains applicable even at long times.   

Our method of noise suppression by  decomposing the displacements into a non-interacting contribution and a collision-induced one is not restricted to hard-spheres 
(see Supplementary Information~\cite{supplement_colloidal_vacf} for simulation results for soft spheres) and should apply also in the case of confinement such as porous media~\cite{Krakoviack:SoftMatter_13:2017,*Krakoviack:PRE_82:2010,*Krakoviack:PRL_94_2005,Pellicane:PRE_69:2004}. Similarly, the strategy should be 
applicable also beyond the mean-square displacement  upon introducing covariances of fluctuating intermediate scattering functions, similar to the measures of dynamic heterogeneities in glassy relaxations~\cite{Glotzer:JCP_112:2000,Flenner:PRL_105:2010,*Flenner:PRE_83:2011,Bertier:Dynamical_Heterogeneities,Avila:PRL_113:2014}. Then, subtle dynamical correlations in Brownian systems that are usually covered by the random fluctuations of the non-interacting system become accessible in simulations. Finally, non-equilibrium simulation set-ups to probe the nonlinear response should also be feasible with the noise-suppression algorithm.

\begin{acknowledgments}
We gratefully acknowledge  Thomas Voigtmann for inspiring discussions  and Rolf Schilling for a critical reading of the manuscript.   
This work has been supported by the Deutsche Forschungsgemeinschaft DFG via the  Research Unit FOR1394 ``Nonlinear Response to
Probe Vitrification'' and the project LO 418/23-1 and  by the Austrian Science Fund
(FWF): I 2887. 
\end{acknowledgments}

%

\onecolumngrid

\section{Supplemental Material}

\section*{1. Noise suppression in the velocity-autocorrelation function}

\begin{figure}[h]
	\includegraphics*[width=.5\linewidth]{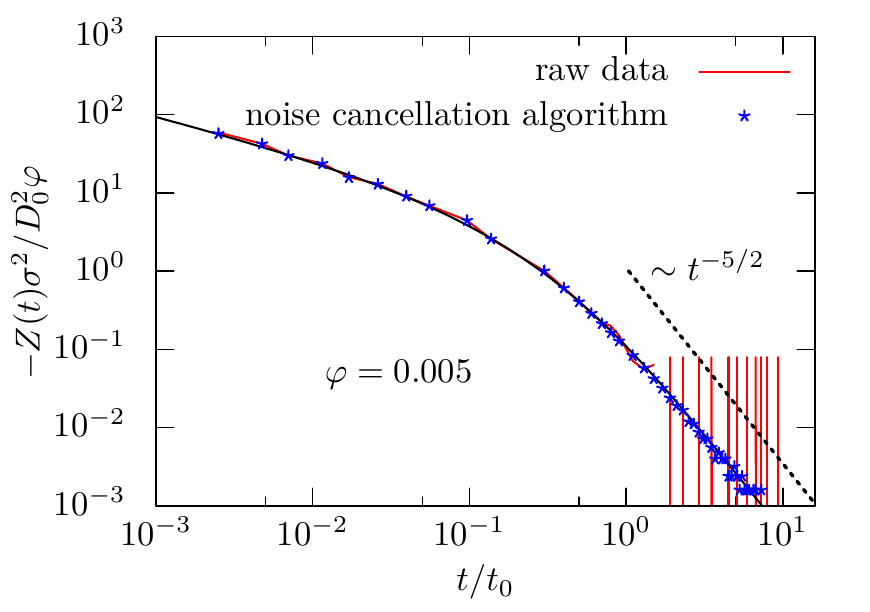}
	\caption{Log-log plot of the reduced VACF at $\varphi=0.005$. The red solid line indicates the VACF obtained from the raw MSD data, whereas blue symbols are obtained using our noise cancellation algorithm. The black solid line is the exact low-density expansion. The dashed black line labeled $t^{-5/2}$ serves as a guide to the eye.}
	\label{fig:benchmark}
\end{figure}

We have argued in the main text that for dilute systems the mean-square displacement (MSD) is dominated by the white noise rather than the interactions. Therefore, we have introduced a novel algorithm which is sensitive only to the collisions. We benchmark our algorithm in Fig.~\ref{fig:benchmark}. The red line indicates the VACF obtained from the raw MSD data, which is too noisy to estimate any signal. By using our noise cancellation algorithm, we enhance the signal-to-noise ratio by two orders of magnitude (blue symbols).

\section*{2. Finite-size effects}

\begin{figure}[h]
	\includegraphics*[width=.6\linewidth]{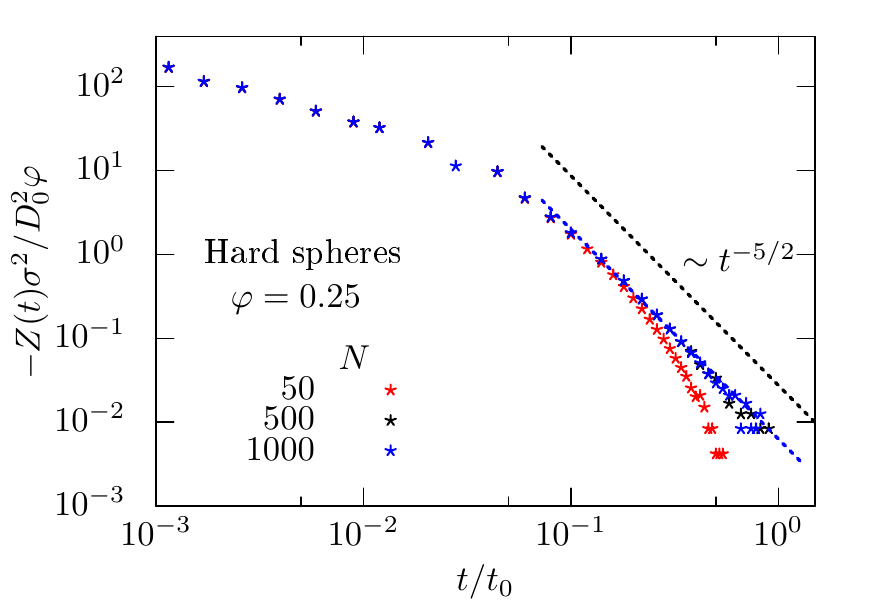}
	\caption{The time dependence of the VACF for all system sizes investigated at a fixed packing fraction $\varphi=0.25$. The dashed blue line is a fit to the power-law tail. The dashed black line labeled $t^{-5/2}$ is added as guide to the eye.}
	\label{fig:finite_size_effects}
\end{figure}

To investigate the finite-size effects on the VACF, we perform Brownian dynamics simulations for the systems of $50$, $500$, and $1000$ hard spheres at a fixed packing fraction $\varphi=0.25$, see Fig.~\ref{fig:finite_size_effects}. The results indicate that finite-size effects exist for the smallest system size (50 particles) and affect the long-time tail. As soon as the particle numbers are larger than 500, we do not observe any finite-size effects. We also repeat the same analysis for other packing fractions, and find similar behavior.

\section*{3. Logarithmic grid for MCT calculation}
For the numerical solution of the MCT, Eqs.~(6) and~(7) in the main text, the intermediate scattering function of the collective $S(k,t)$ and self-motion $S^{(s)}(k,t)$ are required as input. Therefore, the corresponding standard MCT equations~\cite{Goetze:Complex_Dynamics} have to be solved first. For instance, the exact equation of motion for the collective intermediate scattering function in terms of the irreducible memory kernel $m(q,t)$ is given by
\begin{align}\label{eq:MCT_eom}
\tau(q)\dot{S}(q,t)+S(q,t)+\int_0^t m(q,t-t')\dot{S}(q,t')\diff t'=0,
\end{align}
with $\tau(q)=S(q)/q^2 D_0$. The initial condition for the intermediate scattering function is $S(q,t=0) = S(q)$, with $S(q)$ the structure factor.  Within MCT the memory kernel $m(q,t)$ is a functional of the intermediate scattering functions itself, hence Eq.~\eqref{eq:MCT_eom} has to be solved self-consistently with the memory kernel. 
Introducing bipolar coordinates, the standard expression  for the memory kernel reduces to a twofold integral~\cite{Goetze:Complex_Dynamics} 
\begin{align}\label{eq:kernel_collective_scattering}
m(q,t)= \frac{nS(q)}{32\pi^2 q^5}\int_0^\infty k \diff k \int_{|q-k|}^{q+k}p\diff p\big[(q^2+k^2-p^2)c(k)+(q^2+p^2-k^2)c(p)\big]^2S(k,t)S(p,t).
\end{align}
A convenient procedure to reduce the computational complexity is the separation into $p$ and $k$ dependent terms such that the inner integral only depends on $k$ due to the integral bounds $|q-k|$ and $q+k$. For equidistant grids in wave number space the evaluation of the related integral is straightforward.

We, however, rely on a wave vector grid with logarithmic spacing in the reduced wave number, $[x^{-N}, \ldots, x, 1] k_{\text{max}} \sigma$. Then, the boundaries of integrals do not coincide with the wave number 
grid and have to be interpolated to the nearest grid points. Thus, multiple terms have to be considered during the recursive calculation of this integral, in contrast to the equidistant case where only one 
term is relevant in each iteration step. By this adaption to  the logarithmic spacing, the efficient MCT algorithm can be reused and very low wave numbers can be covered. 

To avoid cancellation errors in the mode-coupling functional a Taylor expansion of $\int_{|q-k|}^{q+k}\diff p \dots$ for small wave numbers has been performed in Eq.~\eqref{eq:kernel_collective_scattering}.

A high wave number cutoff of $k_\text{max}\sigma  = 160$  is chosen,  which is considerably higher than in standard MCT discretization. Together with the base $x = 1.02$ and $N=800$ grid points a minimal wave number $k_\text{min} \sigma = x^{-N} k_\text{max} \sigma  = 2.2 \times 10^{-5}$ is reached while the static structure factor is still properly resolved. More grid points have only a marginal effect on the numerical solution.

\section*{4. Closure relations} 
For the structure factors and direct correlation functions we use the Percus-Yevick (PY) closure, which is quite accurate for hard-sphere systems~\cite{Hansen:Theory_of_Simple_Liquids}. For comparison, we also employed the Rogers-Young (RY) closure relation~\cite{Rogers:PhysRevA_30:1984, Heinen:JCP_134:2011,Heinen:JCC_35:2014} which yields thermodynamically consistent results.

\begin{figure}[h]
	\includegraphics*[width=.6\linewidth]{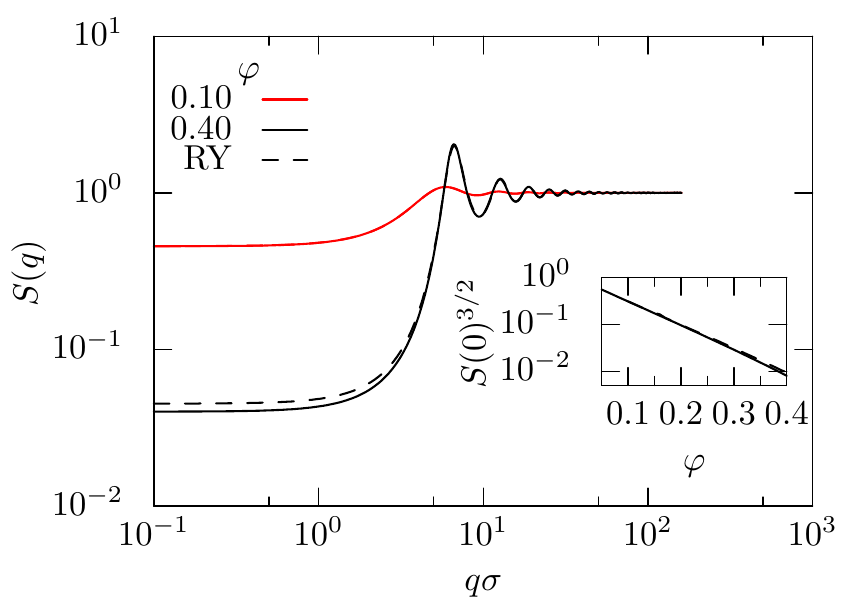}
	\caption{Static structure factor at $\varphi=0.10$ and $\varphi=0.48$ for Percus-Yevick (PY) closure (solid lines) and Rogers-Young (RY) closure (dashed lines). For rather low packing fractions ($\varphi=0.10$) the structure factor for PY and RY closure lie on top of each other. Inset: $S(0)^{3/2}$ vs. packing fraction both for PY and RY.}
	\label{fig:structure_factor_closure}
\end{figure}

For rather low packing fractions ($\varphi=0.10$) PY and RY basically yield the same results (difference less than 1\%), see Fig.~\ref{fig:structure_factor_closure}. For higher packing fractions ($\varphi=0.40$) the absolute value of the structure factor in the long wavelength limit for RY closure is approximately 10\% higher than for PY. Nevertheless, the differences between the closures are insignificant for the dramatic suppression of the long-time tail. Thus, the numerical results for the reduced VACF would be virtually on top of each other, see Fig.~\ref{fig:short_time_divergence}. Furthermore, the dominating $S(0)^{3/2}$ dependence of the prefactor of Eq.~(8) in the main text is only marginally affected by the closure, see Fig.~\ref{fig:structure_factor_closure}~(Inset).

\section*{5. Short-time divergence of the VACF} 

\begin{figure}[h]
	\begin{minipage}[t]{0.48\linewidth}
		\centering
		\includegraphics*[width=\linewidth]{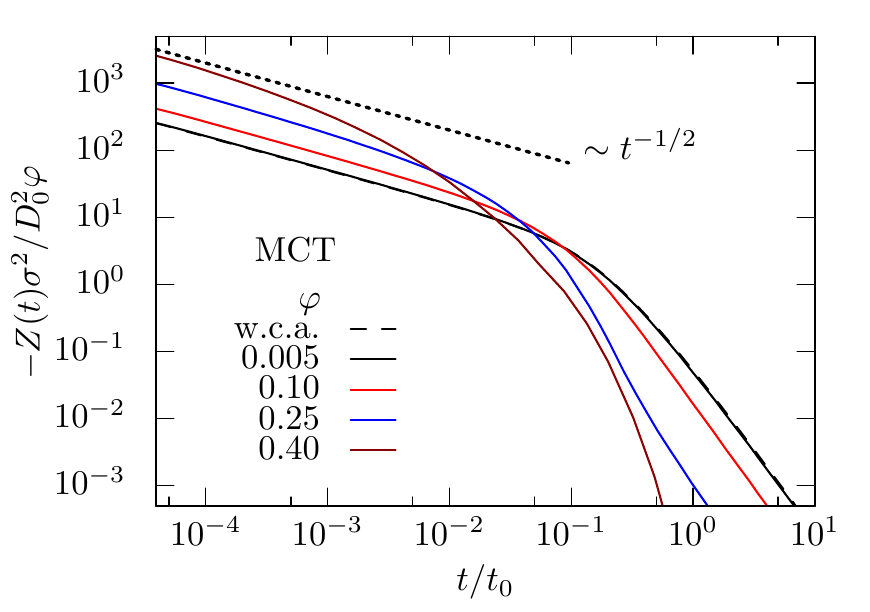}
	\end{minipage}%
	\begin{minipage}[t]{0.48\linewidth}
		\centering
		\includegraphics*[width=\linewidth]{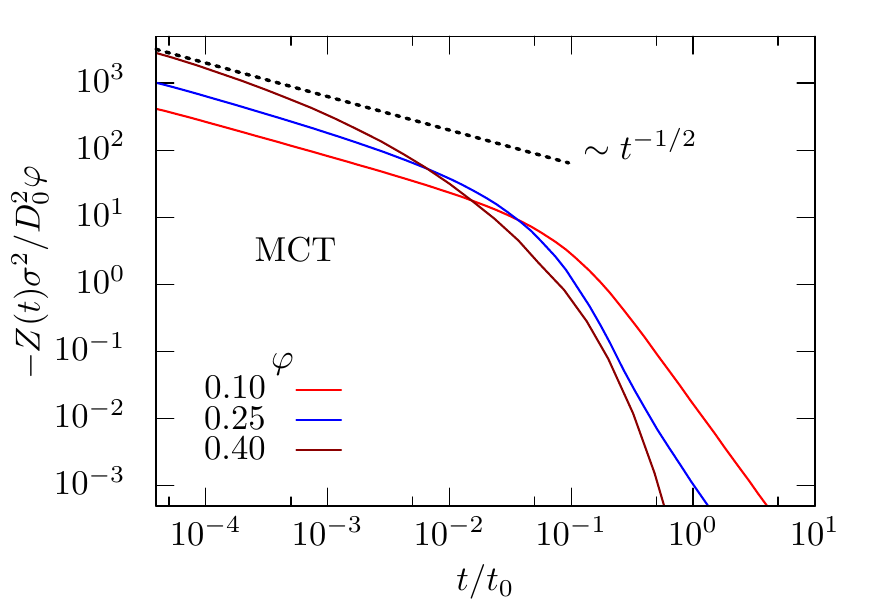} 
	\end{minipage}
	\caption{Mode-coupling theory results for the reduced VACF $Z(t)/\varphi$ using the PY closure in (a) and the RY closure in (b). The short-dashed line indicates a power law $t^{-1/2}$ and serves as a guide to the eye. The long-dashed line in (a) corresponds to the weak-coupling approximation (w.c.a.) and is essentially on top of the numerical MCT results for $\varphi=0.005$.}
	\label{fig:short_time_divergence}
\end{figure}

All simulation data also display a divergent short-time behavior  $Z(t) \sim - t^{-1/2}, t\to 0$ which is a peculiarity of the hard-sphere interaction~\cite{Cichocki:PhysA_204:1994,Banchio:JCP_113:2000}. In fact, for smooth potentials the VACF can be expanded in a regular Taylor series such that the coefficients are determined by static properties~\cite{Tough:MolPhys_1986}. Nevertheless, our data are in agreement with the low-density result $Z(t) \simeq - (8/\sqrt{2\pi}) \varphi (D_0^2/\sigma^2) (t/t_0)^{-1/2}$ for $\varphi\lesssim 0.01$.     

The short-time divergence $t^{-1/2}$ can also be rationalized analytically from the force kernel $\zeta^{(s)}(t)$, Eq.~(7) in the main text. For $t=0$ the integral evaluates to infinity 
due to the exact asymptotic behavior  
of the direct correlation function $n c(k) \sim \cos(k \sigma) / ( k \sigma)^2 $ for $k\to 0$ in the case of  hard spheres (irrespective of the closure relation). 
Yet, for small  $t>0$, the integral converges by the exponential decay of the intermediate scattering functions  $S(k,t) \mapsto S(k) \exp(- D_0 k^2 t/S(k))$ in the short-time diffusive regime. The integral is then dominated by large wave numbers $1 \ll k \sigma \lesssim \sigma/\sqrt{D_0 t} = \sqrt{t_0/t}$, where the integrand averages to a constant, thus yielding $\zeta^{(s)} \sim t^{-1/2}$. 
Then the equation of motion, Eq.~(6) in the main text yields $Z(t) \simeq - D_0^2 \beta^2 \zeta^{(s)}(t) \sim t^{-1/2}$ for $t\to 0$. Our numerical solutions coincide with this predictions for times such that the cut-off in the discretization is irrelevant, $k_\text{max} \sigma \gtrsim \sqrt{t_0/t}$, see Fig.~\ref{fig:short_time_divergence}. To resolve the short-time divergence properly for all packing fractions the time frame starts at $t/t_0=3.9\times 10^{-5}$ corresponding to a high-wave number cutoff $k_\text{max}\sigma  = 160$.

The short-time divergence of the numerical MCT solution coincides with the weak-coupling approximation for $\varphi \leq 0.005$. Interestingly, for the low packing fraction $\varphi = 0.10$, MCT yields already a short-time divergence increased by 60\%, while in our simulations, $Z(t)/\varphi$ is increased by 30\%. The short-time singularity does not arise for smooth potentials, since then the direct correlation functions decay faster, correspondingly, this effect should be viewed as an artifact of the hard-sphere model.

\section*{6. Smooth potentials}

We further perform Brownian dynamics simulations of soft spheres subject to periodic boundary conditions and with Weeks-Chandler-Andersen (WCA) interactions~\cite{Weeks:1971} between particles. These particles have identical diameter $\sigma$ and short-time diffusion coefficient $D_0$. Accordingly, their positions $\vec{R}_i$ evolve as
\begin{equation}
\frac{\diff \vec{R}_i(t)}{\diff t}=- \frac{D_0}{k_BT} \vec{\nabla}_i U_\text{WCA}(R_{ij}) + \sqrt{2D_0} \boldsymbol{\eta}_i(t),
\label{eq:softBD}
\end{equation}   
where $\boldsymbol{\eta}_i$ represents Gaussian white noise with zero mean and unit variance. The distance between particles $i$ and $j$ is denoted by $R_{ij} = |\vec{R}_i -\vec{R}_j|$. The WCA potential is given by
\begin{align*}
U_\text{WCA}(r)=
\begin{cases}
4\epsilon \big [ (\sigma/r)^{12} - (\sigma/r)^{6} \big ] + \epsilon & \text{for } r/\sigma \le 2^{1/6}\\
\quad  0 & \text{else}.
\end{cases}
\end{align*}
Here, $\epsilon>0$ sets the energy scale  and $\sigma> 0$ the length scale. The packing fraction is defined as $\varphi = (N/V)\pi \sigma^3 /6$ similar to the hard-sphere case.   We integrate Eq.~\eqref{eq:softBD} for the temperature $T = \epsilon/k_B$ using the Euler method with a finite time step $\delta t=10^{-5} t_0$, where $t_0=\sigma^2/D_0$ sets the unit of time scale.

\begin{figure}[h]
	\includegraphics*[width=.6\linewidth]{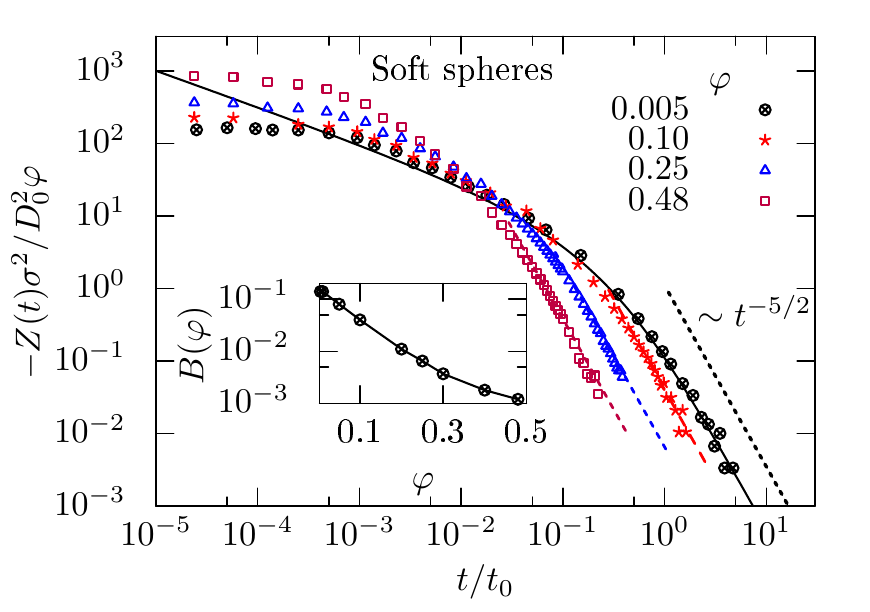}
	\caption{Log-log plot of the reduced VACF of monodisperse Brownian soft spheres for increasing packing fraction $\varphi$. The black is the exact low-density expansion, Eq. (5) in the main text. The colored dashed lines are fit to the data. The dashed black line labeled $t^{-5/2}$ is added as guide to the eye. Inset: Prefactor of the tail vs. packing fraction.}
	\label{fig:softspheres}
\end{figure}

Figure~\ref{fig:softspheres} shows the measured VACF for soft sphere systems using our noise suppression algorithm. The results clearly indicate an anticorrelated long-time tail $Z(t) \simeq - \varphi B(\varphi)(t/t_0)^{-5/2}$ for all packing fractions with dimensionless prefactor $\varphi B(\varphi)$. The prefactor of the long-time tail for various packing fractions is shown in the inset of Fig.~\ref{fig:softspheres}. It turns out that the prefactor is identical to hard sphere systems. Nevertheless, at short times the VACF does not display a divergent behavior, instead it saturates to a finite value. In fact, this peculiarity can be understood by a short-time expansion of correlation functions for smooth potentials, which yields 
\begin{equation}
\langle \Delta \vec{R}(t)^2 \rangle = 6D_0t - \frac{3t^2D_0^2}{k_BT} \big \langle  U_{\text{WCA}}''(R_{ij}) \big \rangle + {\cal O}(t^3),
\label{eq:short-time expansion}
\end{equation}   
where $ \big \langle U_{\text{WCA}}''(R_{ij}) \big \rangle$ evaluates to a positive constant, see Ref.~\cite{Tough:MolPhys_1986} for details. Plugging the above expression into Eq.(1) in the main text, we obtain a finite limiting value $Z(t\to 0) = -(6D_0^2/k_BT) \big \langle U_{\text{WCA}}''(R_{ij}) \big \rangle$ at short times.  

\section*{7. Compressibility effects}
\begin{figure}[h]
	\includegraphics*[width=.8\linewidth]{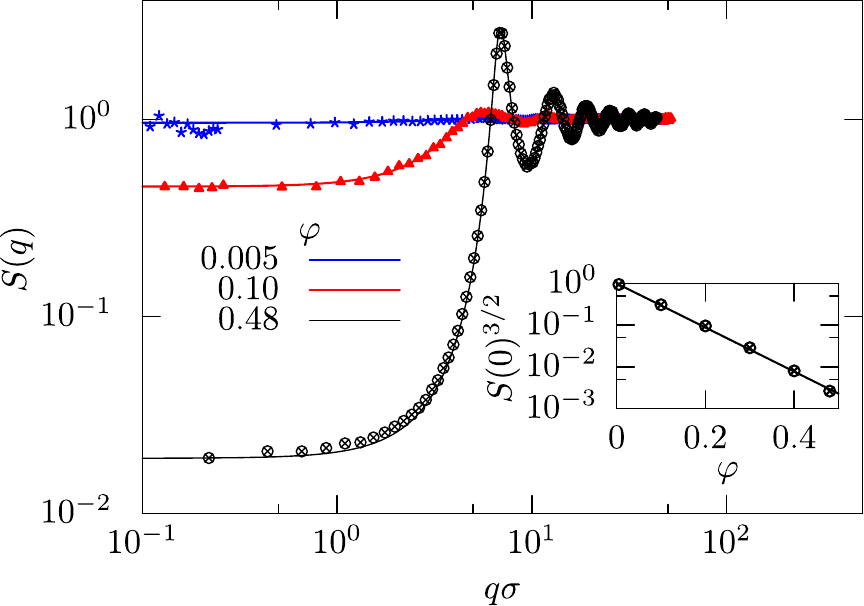}
	\caption{Static structure factor at $\varphi=0.005, 0.10$ and $0.48$. Symbols are from 3d Brownian dynamics simulations and lines are from  PY theory. The inset shows the strong density dependence of the compressibility effects.}
	\label{fig:compressibility}
\end{figure}

To approximate hard spheres interactions, we have also tuned the WCA potential by changing the pair of exponents from $(12-6)$ to $(48-24)$~\cite{Bollinger:SoftMatter2016}. Using this steep potential, we perform Brownian dynamics simulations of 1.3 million particles to extract the density-dependence of the compressibility effects. In Fig.~\ref{fig:compressibility}, we compared the structure factors $S(q)=  |\rho(\vec{q})|^2/N$ from our simulations (symbols) with the PY theory (lines) and found an excellent agreement. In particular, we find that $S(0)^{3/2}$ is suppressed by a factor of 350 with increasing density (see the inset of Fig.~\ref{fig:compressibility}), which suggests that the prefactor of the VACF becomes more dominated by the compressibility effects.

\end{document}